\documentclass[fleqn,10pt]{wlscirep}
\usepackage[utf8]{inputenc}
\usepackage[T1]{fontenc}
\usepackage{aas_macros}
\DeclareUnicodeCharacter{0308}{}

\title{A fast deep-learning approach to probing primordial black hole populations in gravitational wave events}

\author[1]{Jun-Qian Jiang}
\author[1,2]{Hai-Long Huang}
\author[3]{Jibin He}
\author[4,5,1,2]{Yu-Tong Wang}
\author[1,2,4,6]{Yun-Song Piao}
\affil[1]{School of Physical Sciences, University of Chinese
Academy of Sciences, Beijing 100049, China} \affil[2]{School of
Fundamental Physics and Mathematical
    Sciences, Hangzhou Institute for Advanced Study, UCAS, Hangzhou
    310024, China}
\affil[3]{Department of Physics and Chongqing Key Laboratory for
Strongly Coupled Physics, Chongqing University, Chongqing 401331,
P. R. China} \affil[4]{International Center for Theoretical
Physics
    Asia-Pacific, Beijing/Hangzhou, China}
\affil[5]{Taiji Laboratory for Gravitational Wave Universe (Beijing/Hangzhou), UCAS, Beijing 100190, China}
\affil[6]{Institute of Theoretical Physics, Chinese
    Academy of Sciences, P.O. Box 2735, Beijing 100190, China}%

\begin{abstract}
Primordial black holes (PBHs), envisioned as a compelling dark
matter candidate and a window onto early-Universe physics, may
contribute to some of the gravitational-wave (GW) signals
detected by the LIGO-Virgo-KAGRA network. Traditional hierarchical
Bayesian analysis, which relies on precise GW-event posterior
estimates to extract information on potential PBH populations from
GW events, becomes computationally demanding for catalogs with a large number of events.
Here, we present a fast
deep-learning framework, leveraging Transformer and normalizing
flows, that maps GW-event posterior samples to joint posterior
distributions over the hyperparameters of the PBH population.
Our approach yields credible intervals with acceptable accuracy while delivering an order-of-magnitude speedup.
These results highlight the potential of deep learning for fast and accurate PBH population studies, and its applicability to next-generation GW detectors when combined with appropriate event-level inference models.

\end{abstract}
\begin{document}

\flushbottom
\maketitle
\thispagestyle{empty}

\maketitle

The fast and accurate identification of potential PBH populations
in GW events is an important challenge. It has been widely thought
that PBHs, first proposed as relics of the early Universe
\cite{Hawking:1971ei,Carr:1974nx,Zeldovich:1967lct}, can be a
tantalizing candidate for (at least part of) the dark matter
budget~\cite{Carr:2016drx,Chapline:1975ojl,Meszaros:1975ef,Carr:2020gox,Calza:2024fzo,Calza:2024xdh}
and possible seeds for the supermassive black holes in galactic
nuclei~\cite{Carr:2023tpt,Nakama:2016kfq,Hai-LongHuang:2024vvz,Hai-LongHuang:2024gtx}.
Yet, despite decades of well-motivated models~\cite{Carr:1993aq,Ivanov:1994pa,Garcia-Bellido:1996mdl,Randall:1995dj} and
indirect bounds for PBHs~\cite{Murgia:2019duy,Bird:2016dcv,Sasaki:2016jop}
(see also \cite{Sasaki:2018dmp,Domenech:2024cjn} for recent reviews), whether they exist or not
has remained elusive. Recent GW observations open a new window:
mergers of PBH binaries would exhibit population-level signatures,
distinct mass spectra~\cite{He:2023yvl,Andres-Carcasona:2024wqk},
redshift evolution of the merger
rate~\cite{Stasenko:2024pzd,Raidal:2024bmm}, spin
distributions~\cite{DeLuca:2023bcr,DeLuca:2020bjf}, and even
clustering
properties~\cite{Huang:2023mwy,Crescimbeni:2025ywm,Clesse:2024epo}
that differ markedly from those of astrophysical black holes
(ABHs). How to fully utilize the GW observations to search for
PBHs in a black hole population composed of PBHs and ABHs is
crucial for enhancing our understanding of PBHs and the early
Universe (see e.g.~\cite{LISACosmologyWorkingGroup:2023njw} for a
recent review).

{
Hierarchical Bayesian inference (HBA) is the standard tool for extracting population hyperparameters, such as the primordial black hole abundance $f_\text{PBH}$, from GW catalogs
} \cite{He:2023yvl,Huang:2024wse,Andres-Carcasona:2024wqk},
relying critically on accurate single-event posterior samples.
However, as detections climb into the hundreds, the combined cost
of generating per-event posteriors and sampling the population
likelihood grows rapidly, rendering full hierarchical analysis
increasingly challenging.
Neural network based single-event inference schemes
(e.g.~\cite{Leyde:2023iof}) can accelerate the former process, but inferring black hole population properties from event-parameter posterior distributions remains a bottleneck.
Given the large number of events expected from
upcoming detector networks, there is a growing need for new
methods to efficiently measure PBH population hyperparameters from
GW events.

Deep learning based inference may provide a viable alternative approach.
For example, Ref.~\cite{Wong:2020jdt} applied it to population parameter inference in the absence of measurement systematics and statistical uncertainties, while Ref.~\cite{Cheung:2021orb} used the neural network as an emulator for the population probability density function.
Here we present a deep learning based inference approach, built
upon Transformer architectures and normalizing flows, that maps collections of GW posterior samples onto joint PBH
population posteriors, to overcome these challenges.
{
Our method delivers estimates of key hyperparameters that are consistent with those obtained from HBA within the same inference framework, while achieving an order-of-magnitude speedup.
}
As a result, it provides a fast alternative with acceptable accuracy.

\section*{Results}

\subsection*{Deep learning network}

\begin{figure}[!htb]
    \centering
   \includegraphics[width=1\textwidth]{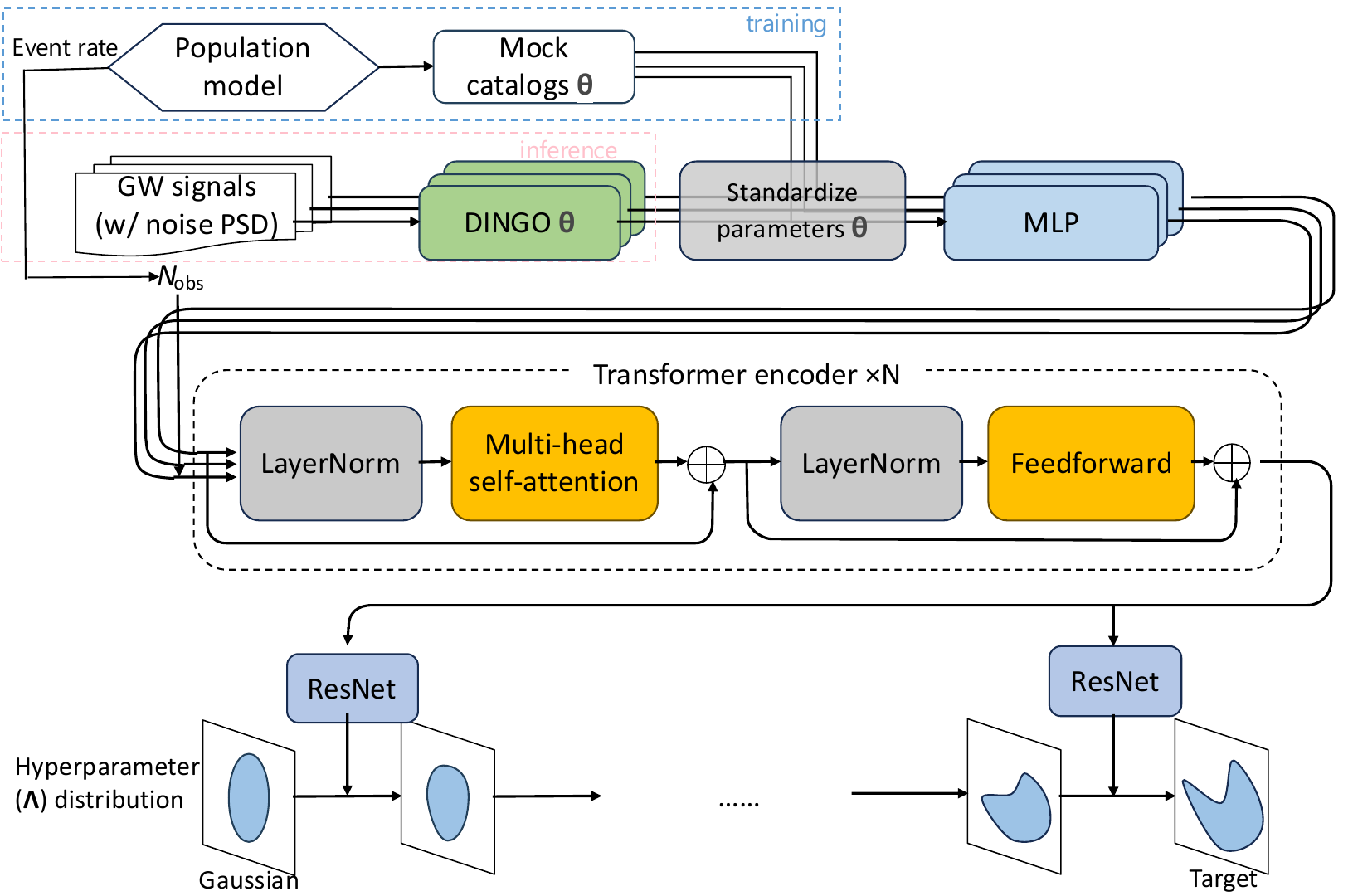}
    \caption{\textbf{The architecture of our deep learning framework.}
During training, we generate event catalogs based on the assumed black hole population model; during inference, we use DINGO~\cite{Leyde:2023iof} to infer the posterior distribution of parameters $\theta$ for each GW event and construct the catalog accordingly.
The resulting posterior samples are processed through a shared
multilayer perceptron (MLP), aggregated together, and combined
with the observed number of events $N_\text{obs}$. A Transformer encoder is
then applied to capture the population-level features of the black
holes.
{
Finally, a normalizing flow is used to model the joint posterior distribution over the population hyperparameters $\Lambda$, transforming a simple Gaussian distribution into the target distribution via a sequence of invertible transformations parameterized by residual networks (ResNets). Details of the network architectures are provided in Sec. \textbf{Methods}.
}
}
    \label{fig:CNF}
\end{figure}

Our deep learning approach consists of four steps, as illustrated
in \autoref{fig:CNF}.
{
During training, we first generate large event catalogs based on the assumed black hole population model, where each catalog is characterized by a lot of event-level parameters $\mathbf{\theta}$ and the observable event rate.
We then select the same number of events as in the real inference setup for training (See \textbf{Generating the data set} of \textbf{Methods} for details).
During inference, starting from GW waveforms, we use DINGO~\cite{Leyde:2023iof}, a machine learning based fast parameter inference tool, to analyze each event together with the corresponding noise power spectral density (PSD) and infer the event parameters.
}
In this work, we aim to extract information
about the black hole population model from the mass and redshift
distributions. Therefore, the event-level parameters of interest,
$\mathbf{\theta}$, are the component black hole masses $m_1,\,
m_2$ and the luminosity distance $d_L$.
Next, we draw 10,000 posterior samples for each event, ensuring that the effective sample size~\cite{Farr:2019rap,Talbot:2023pex} in the likelihood evaluation (\autoref{eq:event_MCint}) exceeds $\sim 1000$.
After standardizing each parameter individually (see \textbf{Neural network architecture} of \textbf{Methods} for details), we use a shared
MLP to summarize the distribution into 31
scalars for each event. By combining these compressed
representations with the observed number of events $N_\text{obs}$, we
obtain a sequence whose length corresponds to the number of
observed events. We use a Transformer encoder to capture
information about the black hole population model encoded in the
distribution of $\theta$. To obtain the joint distribution over
the population hyperparameters $\Lambda$, we use a normalizing
flow that progressively transforms a Gaussian distribution into
the target distribution, with each transformation step governed by
a residual network (ResNet). The details of these networks are
provided in Sec. \textbf{Methods}. The DINGO model used here is
pre-trained, while the remaining neural networks are jointly
trained.

\subsection*{Parameter inference for PBHs in GW events}
\label{sec:Results}

As a proof of concept, we consider a black hole population model
composed of PBHs and ABHs. In the model we investigate, PBHs are
from the post-inflation collapses of supercritical bubbles during
inflation and are described by three parameters: the
characteristic mass $M_c$ of the black hole mass distribution in
their source frame, the width of the mass distribution $\sigma$,
and the fraction of PBHs in dark matter $f_\text{PBH}$~\cite{Hai-LongHuang:2023atg,Liu:2018ess}.
Meanwhile,
ABHs are described by the phenomenological POWER-LAW + PEAK model~\cite{LIGOScientific:2018jsj,DeLuca:2021wjr,KAGRA:2021duu,Madau:2014bja}.
In this work, we focus on the inference of PBH parameters, i.e.
$\{M_c,\,f_\text{PBH},\,\sigma\}$, and consider observations
consisting of 128 merger events, with a threshold of signal-to-noise ratio (SNR) $\gtrsim
12$ assuming an O3 noise curve for observability. The corresponding
details are provided in \textbf{Methods}.

\begin{figure}[!htb]
    \centering
   \includegraphics[width=0.48\textwidth]{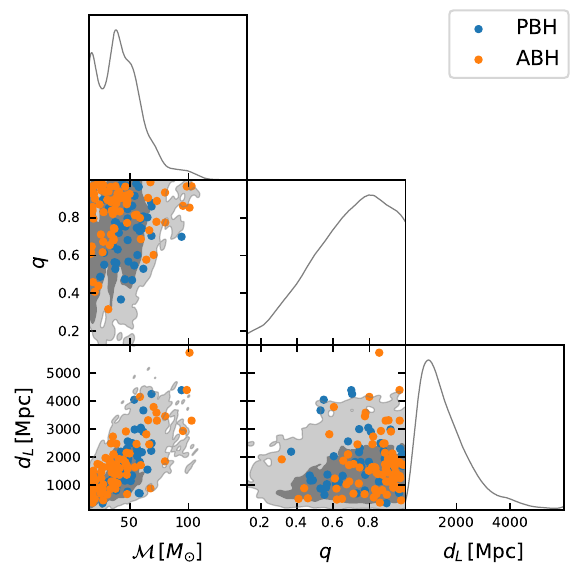}
   \includegraphics[width=0.48\textwidth]{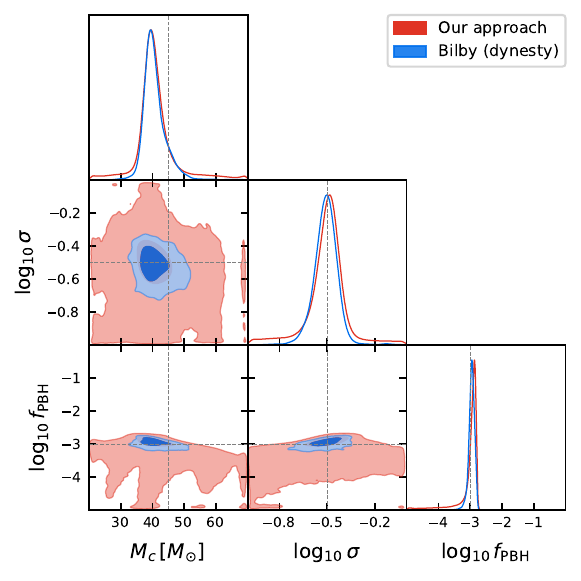}
\caption{\textbf{\textit{Left}: The illustration of the inputs to our model, i.e., mock observation event parameter
distribution.} Gray contours show joint posterior of chirp mass
$\mathcal{M}$, mass ratio $q$, and luminosity distance $d_L$
inferred by DINGO for all events, which serve as input to
our population analysis.
    Blue points mark the true parameters of individual PBH merger events, while orange points mark the true parameters of individual ABH merger events.
    \textbf{\textit{Right}: Hyperparameter inference for PBHs in
a black hole population model composed of PBHs and ABHs
($f_\text{PBH} = 10^{-3}$).}
    The red regions show the inference results from our approach, while the blue regions show the results from Hierarchical Bayesian inference using dynesty for sampling.
    The injected values are indicated by dashed lines.
    In both panels, the contours on the two-dimensional plane represent the 68\% and 95\% confidence regions.}
    \label{fig:dist_fPBH-3}
\end{figure}

{
To investigate the robustness of our model, we fix
$f_{\rm PBH} = 10^{-3}$ and set other values compatible with current observations.
\footnote{We fix $\log_{10} R^0_\text{ABH} = \log_{10} 20,\, \log_{10} \lambda = \log_{10} 0.038,\, m_\text{min} = 10,\, m_\text{max} = 80,\, \alpha = 3.5,\, \mu_G = 35,\, \sigma_G = 3,\, \delta_m = 6,\, \beta = 1.2,\, M_c = 45,\, \log_{10} \sigma = -0.5$.}
Out of the 128 generated events, 46 are PBH binaries.}
{
We show the illustration of the inputs to our model in the left panel of \autoref{fig:dist_fPBH-3}, while we show the true parameters of each event as colored points and overlay the posterior distributions of all events inferred by DINGO in gray.
These posteriors are based on the parameter priors used in DINGO training, and this prior dependence is removed through reweighting (\autoref{eq:event_MCint}).
}
Traditional
hierarchical Bayesian inference on the PBH population parameters
using the dynamic nested sampling package \texttt{dynesty}~\cite{2020MNRAS.493.3132S} (wrapped in \texttt{bilby}~\cite{Ashton:2018jfp}) has also been performed,
and the comparison of the results between hierarchical Bayesian
inference and our deep-learning approach is presented in the right
panel of \autoref{fig:dist_fPBH-3}.

For all sampled PBH population hyperparameters, $M_c$, $\sigma$, and $f_\text{PBH}$, our approach produces posterior distributions whose peaks are consistent with those from \texttt{dynesty}.
In addition, for the marginalized one-dimensional posterior distributions, our results are very consistent with those from \texttt{dynesty}.
The same behavior is observed in the two-dimensional parameter space, but with larger uncertainty for the 95\% credible region.
{
In the full 12-parameter validation example reported in the Supplementary material, we generated \(6,080,000\) samples from the learned posterior and applied likelihood-based reweighting using the HBA likelihood. The resulting effective sample size is \(N_{\rm eff}=469,351\), indicating that, in this example, the learned posterior has adequate support for likelihood-based post-processing and that the reweighted result is not dominated by a small number of samples.
}

{
\subsection*{Speed comparison}

We measured runtime on a compute node equipped with two Intel Xeon Gold 6240R CPUs (48 cores in total) and four NVIDIA A100 GPUs.

For the traditional method \texttt{dynesty}, we benchmarked both CPU and GPU implementations: the former based on NumPy and the latter based on CuPy (similar to Ref.~\cite{Mastrogiovanni:2023zbw}).
In both cases, we used multiprocessing parallelization (distributed across different CPU cores or GPUs) provided by \texttt{dynesty}, and vectorized the merger-rate calculation over the event-parameter vector $\theta$.
Following the default \texttt{bilby} settings, we used 1000 live points and adopted d$\ln \mathcal{Z} < 0.1$ as the stopping criterion.
As a result, we obtained 8333 posterior samples.
For the CPU-based implementation, the runtime was 187 hours, whereas for the GPU-based implementation it was 63 hours.
In addition, we tested Metropolis-Hastings Markov Chain Monte Carlo sampling, but found that it would require well over 7 days.
Therefore, we did not complete that run.

For our deep-learning approach, the full pipeline consists of three stages: data generation, neural network training, and inference.
In our experiments, data generation took 4 hours.
However, this stage only involves generating a large number of event parameters $\theta$ and the expected detectable event rate, which means that as long as the detectability threshold remains unchanged, the generated training set can be reused for future observations over longer durations.
Moreover, because each mock sample is independent, the workflow is naturally scalable and can be readily parallelized across additional computational resources.
Neural network training took 1 hour.
Finally, the inference time depends on the number of posterior samples required. In practice, $<10$ s is usually sufficient, so this cost is negligible.

Overall, the deep-learning approach has an order-of-magnitude advantage in runtime over traditional inference methods.

}

\subsection*{Validation of inference results}

\begin{figure}[!htb]
    \centering
   \includegraphics[width=0.7\textwidth]{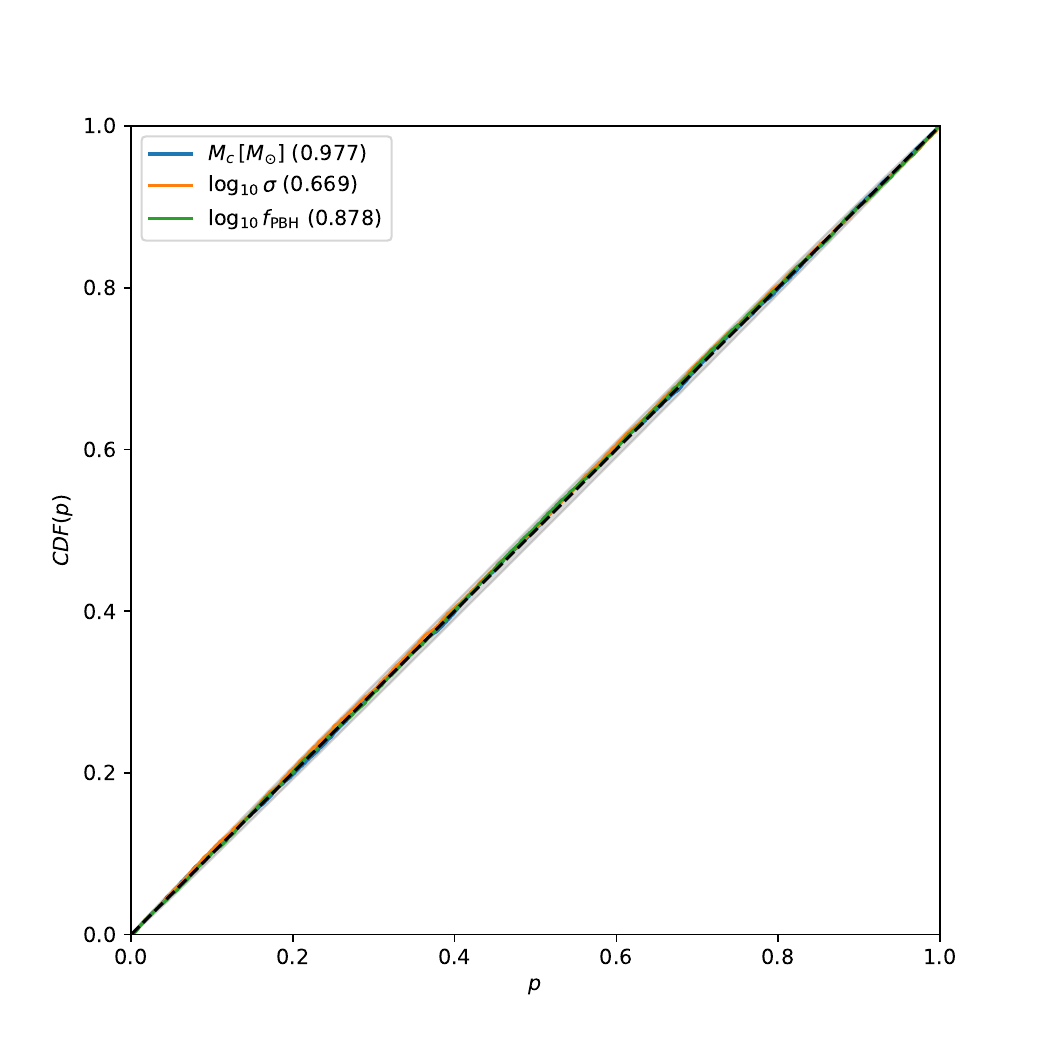}
\caption{\textbf{Probability-Probability plot for a set of 10000
posterior evaluations.} Each cumulative distribution aligns well
with the diagonal, with the spread mostly confined within the
$2\sigma$ gray regions for almost the entire confidence level
interval. The legend shows the $p$-values of the individual
parameters.}
    \label{fig:pp_plot}
\end{figure}

{To test the marginal calibration of the learned posterior}, we randomly draw posterior samples
from 10,000 model realizations and construct a
Probability-Probability plot shown in Fig.~\ref{fig:pp_plot}.
For
each hyperparameter, we compute the percentile score of the true
value within its marginalized posterior, and plot the
corresponding cumulative distribution function (CDF). For the true
posteriors, the percentiles should be uniformly distributed,
meaning the CDF should follow a diagonal line. The
Kolmogorov-Smirnov test $p$-values are reported in the legend,
ranging from $66.8\%$ to $97.7\%$.
{
For completeness, we also
combine them using Fisher's method,
$
    p_{\rm comb}
    =
    P\left(
    \chi^2_{2k}
    >
    -2\sum_{i=1}^{k}\ln p_i
    \right),
$
with $k=3$, obtaining $p_{\rm comb}=0.981$.
This combined value should
be interpreted only as a compact summary of the marginal calibration
diagnostics. We emphasize that the P--P plots test the calibration
of the one-dimensional marginalized posteriors within the assumed
simulation framework. They do not, by themselves, constitute a
validation of the full joint posterior geometry.
We present several more DL-vs-HBA validation examples in \textbf{Supplementary material}.
}

\section*{Discussion}
\label{sec:conclusions}

It is expected that as next-generation GW observatories begin
operation and detection rates rise, the growing volume of data
will call for more efficient analysis methods. Deep learning
approaches like ours will be essential for drawing out
cosmological and population-level insights from these
observations.
In this work, we have presented a new deep-learning framework that
uses Transformer encoders and normalizing flows to infer the main
parameters of the PBH populations directly from GW catalogs. Our
method achieves an order-of-magnitude speedup over standard
hierarchical Bayesian samplers.

{
One key advantage of our approach is that it eliminates the need
to extensively evaluate an explicit likelihood function, a major
computational bottleneck in standard hierarchical Bayesian
inference.
}
By rapidly estimating the PBH to dark matter fraction
$f_\text{PBH}$ and other population parameters, it enables swift
population studies that can
{
help place constraints on the primordial black hole population and other population models.
}
This capability will
be especially valuable for the large catalogs expected from future
space-based detectors such as LISA~\cite{2017arXiv170200786A},
Taiji~\cite{Hu:2017mde}, Tianqin~\cite{TianQin:2015yph}, as well
as next-generation ground-based observatories. Such computational
efficiency not only keeps pace with the growing GW event rate but
also allows for rapid updates to cosmological and dark matter
constraints. Moreover, our framework is readily extensible: it can
incorporate more sophisticated population models, such as those
including accretion histories or detailed spin distributions.

{
We emphasize that the present method is designed to emulate the
hierarchical Bayesian population inference defined by a specified
population model and selection function, rather than as a
method that automatically removes systematic errors due to
population model misspecification. 
Similarly, the simplified selection prescription used in
this proof-of-concept study is not intended to represent the full
selection function of a real search pipeline. In realistic
applications, this prescription should be replaced by the same
injection based or search pipeline based selection function used in
standard hierarchical Bayesian population inference.
}

There remain several avenues for further development. In our
current design, the network is trained for a fixed catalog size.
Handling different numbers of events requires either adjusting the
width of the ResNet layers or processing events in batches with
appropriate reweighting.
It is also possible to design alternative sequence models that accommodate variable-length inputs without manual grouping.
Moreover, rather than drawing posterior
samples from DINGO for each event, one could feed the activations
of DINGO’s final hidden layer directly into our framework,
reducing sampling noise and improving end-to-end efficiency. The
fidelity of our results depends on the coverage and realism of the
training set, so extending the parameter space and incorporating
additional astrophysical effects will be important to guard
against out-of-distribution biases. Finally, integrating
multi-messenger data streams and refining the PBH models by
including detailed spin evolution or accretion physics will be
crucial steps toward a comprehensive characterization of the PBH
populations.

\section*{Methods} \label{sec:method}

\subsection*{Black hole binary population model}
\label{sec:BH}

PBHs can form binaries through several formation channels both in the early Universe when two PBHs are produced sufficiently close to each other \cite{Bird:2016dcv,Sasaki:2016jop,Nakamura:1997sm}, and in the late Universe by capture in clusters
\cite{Bird:2016dcv,Nishikawa:2017chy,Ali-Haimoud:2017rtz},
as discussed in the recent review by \cite{Raidal:2024bmm}.
In this study, we focus on the formation channel of PBH binaries in the early
Universe, which is known to make a dominant contribution to the PBH merger
rate \cite{Raidal:2017mfl,Franciolini:2022ewd,Raidal:2024bmm}.
In this case,
the merger rate density per unit volume at cosmic time $t$ for PBHs is
\cite{Hai-LongHuang:2023atg,Liu:2018ess}
\begin{align}\label{eq:merger_rate_PBH}
    \frac{{\rm d}R_{\rm PBH}}{{\rm d}m_i{\rm d}m_j}&\approx\frac{1.99\times10^6}{\text{Gpc}^3\text{yr}}f^{1.46}
    \left(1+\frac{\sigma_{\text{eq}}^2}{f^2}\right)^{-0.27}
    \left(\frac{m_i}{M_\odot}\right)^{-0.92}\left(\frac{m_j}{M_\odot}\right)^{-0.92}
    \notag \\ & \times
    \left(\frac{m_i+m_j}{M_\odot}\right)^{0.97}
    \left(\frac{t}{t_0}\right)^{-0.92}\psi(m_i)\psi(m_j),
\end{align}
where $f\approx0.85f_{\rm PBH}$ is the total abundance of
PBHs in nonrelativistic matter, $t_0$ is the present time and
$\sigma_{\rm eq}^2$ is the variance of density perturbations
of the rest of dark matter at $z_{\rm eq}$.
We will focus only on PBHs in the stellar mass range, thus the effect of cosmic expansion on the comoving distance of PBH pairs is negligible \cite{Hai-LongHuang:2023atg}. In addition, we assume that PBHs are initially randomly distributed according to a spatial Poisson distribution. Generalizing to the case of initial clustering is straightforward \cite{Huang:2023mwy,Desjacques:2018wuu,Inman:2019wvr,DeLuca:2020jug}.

The PBH mass function in \eqref{eq:merger_rate_PBH} is defined by $\psi(m)\equiv\frac{m}{\rho_{\rm PBH}}\frac{{\rm d}n_{\rm PBH}}{{\rm d}m}$.
As an example, we consider a mass function where PBHs are sourced by supercritical bubbles that nucleated during slow-roll inflation
\cite{Huang:2023chx,Huang:2023mwy}
\begin{equation} \label{eq:bubbleMFa2}
    \psi_{\rm bubble}(m|M_c,\sigma)=e^{-\sigma^2/8}\sqrt{\frac{M_c}{2\pi\sigma^2 m^3}}\exp
    \left(-\frac{\ln^2(m/M_c)}{2\sigma^2}\right).
\end{equation}
As mentioned in the introduction, we use the deep learning tool DINGO
\cite{Dax:2021tsq}
to analyze the strain data and generate single-event posterior samples for the subsequent population analysis.

We consider events that fall within the prior range used by the pre-trained DINGO model, namely luminosity distance $d_L \in [100, 6000]$ Mpc\footnote{The conversion between luminosity distance and redshift uses the Planck 2018 results~\cite{Planck:2018vyg}.}, component masses $m_1, m_2 \in [10, 180]$ $M_\odot$, chirp mass $\mathcal{M} \in [10, 180]$ $M_\odot$, and mass ratio $q \in [0.125, 1]$.
{
For the masses, we use the flat prior defined on the $m_1$ - $m_2$ plane, and the chirp mass and mass ratio ranges are imposed as hard boundaries.
For the distance, we use a flat prior defined on the luminosity distance.
}
This same selection is also
made for the case of the ABH model below.

To describe the ABH population, we use the widely used phenomenological POWER-LAW + PEAK model \cite{LIGOScientific:2018jsj,DeLuca:2021wjr,KAGRA:2021duu,Madau:2014bja} as an example to model the differential merger rate of ABHs:
\begin{equation} \label{eq:merger_rate_ABH}
    \frac{{\rm d}R_{\rm ABH}}{{\rm d}m_1{\rm d}m_2}=R_{\rm ABH}^0(1+z)^\kappa
    p_{\rm ABH}^{m_1}(m_1)p_{\rm ABH}^{m_2}(m_2|m_1),
\end{equation}
where $R_{\rm ABH}^0$ is the local merger rate at redshift $z=0$, and $\kappa\simeq2.9$
describes the merger rate evolution with redshift \cite{KAGRA:2021duu,Madau:2014bja}.
The probability density function of the primary mass is modeled as a combination
of a power law and a Gaussian peak
\begin{equation}
    p_{\rm ABH}^{m_1}(m_1)=\left[(1-\lambda)P_{\rm ABH}(m_1)+\lambda G_
    {\rm ABH}(m_1)\right] S(m_1|\delta_m,m_{\rm min}),
\end{equation}
where
\begin{equation}
    P_{\rm ABH}(m_1|\alpha,m_{\rm min},m_{\rm max})\propto\Theta(m-m_{\rm min})
    \Theta(m_{\rm max}-m)m_1^{-\alpha},
\end{equation}
\begin{equation}
    G_{\rm ABH}(m_1|\mu_G,\sigma_G,m_{\rm min},m_{\rm max})\propto\Theta(m-m_{\rm min})
    \Theta(m_{\rm max}-m)\exp\left(-\frac{(m_1-\mu_G)^2}{2\sigma_G^2}\right)
\end{equation}
are restricted to masses between $m_{\rm min}$ and $m_{\rm max}$ and normalized.
The term $S(m_1|m_{\rm min},\delta_m)$ is a smoothing function, which rises from
0 to 1 over the interval $(m_{\rm min}, m_{\rm min}+\delta_m)$,
\begin{equation}
    S(m \mid m_{\rm min}, \delta_m) = \begin{cases}
        0, & m< m_{\rm min} \\
        \left[f(m - m_{\rm min}, \delta_m) + 1\right]^{-1}, & m_{\rm min} \leq m < m_{\rm min}+\delta_m \\
        1, & m\geq m_{\rm min} + \delta_m
    \end{cases}
\end{equation}
with
\begin{equation}
    f(m', \delta_m) = \exp \left(\frac{\delta_m}{m'} + \frac{\delta_m }{m' - \delta_m}\right).
\end{equation}
The distribution of the secondary mass is modelled as a power law,
\begin{equation}
    p_{\rm ABH}^{m_2}(m_2|m_1,\beta,m_{\rm min})\propto\left(\frac{m_2}{m_1}\right)^{\beta}.
\end{equation}
where the normalization ensures that the secondary mass is bounded by
$m_{\rm min}\le m_2\le m_1$.

In this work, we consider black hole binaries (BHBs) with primordial and astrophysical origin (PBH+ABH).
Therefore, the merger rate of ABH is given by \eqref{eq:merger_rate_ABH} and the total merger rate given by ${\rm d}R/{\rm d}m_1{\rm d}m_2={\rm d}R_{\rm PBH}/{\rm d}m_1{\rm d}m_2+{\rm d}R_{\rm ABH}/{\rm d}m_1{\rm d}m_2$.
As a proof of concept, in this work we do not consider mergers between different populations.

\subsection*{Hierarchical Bayesian population methods}
\label{sec:HBA}

The classical approach will function as our reference point against which
we will compare the outcomes with the deep learning methods.
In this section, we introduce the hierarchical Bayesian inference methods, following the approach outlined in \cite{He:2023yvl}.

According to Bayes' theorem, the population posterior distribution
\begin{equation}
    p(\Lambda|\boldsymbol{d})=\frac{\mathcal{L}(\boldsymbol{d}|\Lambda)
    \pi(\Lambda)}{\mathcal{Z}_\Lambda},
\end{equation}
where we define the data measured from observed GW populations as $\boldsymbol{d}$. 
In the hierarchical Bayesian framework, the black hole population model parameters $\Lambda$ that we seek to infer are referred to as hyperparameters, their priors $p(\Lambda)$ are correspondingly called hyperpriors (summarized in \autoref{tab:prior}) and
\begin{equation} \label{eq:evidence}
    \mathcal{Z}_{\Lambda}=\int\mathcal{L}(\boldsymbol{d}|\Lambda)
    \pi(\Lambda){\rm d}\Lambda
\end{equation}
is the hyper-evidence.
The hyper-likelihood is
\cite{Loredo:2004nn,LIGOScientific:2020kqk,2019PASA...36...10T,Mandel:2018mve}
\begin{equation} \label{eq:tot_likelihood}
    \mathcal{L}(\boldsymbol{d}, N_\text{obs}|\Lambda)\propto N(\Lambda)^{N_\text{obs}} e^{-N(\Lambda)\xi(\Lambda)} \prod_{i=1}
    ^{N_{\rm obs}}\int \mathcal{L}(d_i|\boldsymbol{\theta}_i)\pi(\boldsymbol{\theta}_i|\Lambda){\rm d}\boldsymbol{\theta}_i,
\end{equation}
Here, $\mathcal{L}(d_i|\boldsymbol{\theta}_i)$ is the single event likelihood for event parameters $\boldsymbol{\theta}_i$ (We consider their mass and luminosity distance in this work) given the strain data $d_i$.
Their conditional prior $\pi(\boldsymbol{\theta}|\Lambda)$ is given by the population model:
\begin{equation} \label{eq:population_model}
    \pi(\boldsymbol{\theta}|\Lambda) \propto 
    \frac{1}{1+z}\frac{{\rm d}V_c}{{\rm d}z}
    \frac{{\rm d}R}{{\rm d}m_1{\rm d}m_2}(\boldsymbol{\theta}|\Lambda),
\end{equation}
where ${\rm d}V_c/{\rm d}z$ represents the
differential comoving volume.
$N(\Lambda)$ is the expected number of events occurring during the observation time $T_\text{obs}$.
However, not all such events are detectable.
The factor $\xi(\Lambda)$ accounts for the selection biases introduced by the detector’s sensitivity:
\begin{equation}
    \xi(\Lambda)=\int p_{\rm det}(\boldsymbol{\theta})\pi(\boldsymbol{\theta}|\Lambda){\rm d}\boldsymbol{\theta},
\end{equation}
where $p_{\rm det}(\boldsymbol{\theta})$ denotes the detection probability,
and depends primarily on the masses and redshift of the system \cite{LIGOScientific:2020kqk}.
The estimation of $\xi(\Lambda)$ is performed using simulated
injection samples, where
a Monte Carlo integral over $N_{\rm inj}$ injection samples is used to
approximate it as
\begin{equation} \label{eq:injection}
    \xi(\Lambda)\approx\frac{1}{N_{\rm inj}}\sum_{j=1}^{N_{\rm det}}
    \frac{\pi(\boldsymbol{\theta}_j|\Lambda)}{\pi_{\rm inj}(\boldsymbol{\theta}_j)}\equiv\frac{1}
    {N_{\rm inj}}\sum_{j=1}^{N_{\rm det}}s_j,
\end{equation}
where $\pi_{\rm inj}(\boldsymbol{\theta}_j)$ is the prior probability of the $j$-th event
(the probability density function from which the injections are drawn),
$N_{\rm det}$ denoting the count of successfully detected injection samples.
{
The difference between the distribution of injected events and the true distribution can introduce errors in the Monte Carlo estimation when the injection set does not effectively cover the support of the population model.
Therefore, when using an injection set that is not tailored to the target model, it may be necessary to adjust the injections.
In our HBA analysis, we inject $2^{19}$ events under a flat prior on event parameters, then calculate $s_j$ and average it, yielding an effective number of injection samples, $N_{\rm eff}^{\rm inj}=(\sum_j s_j)^2/\sum_j s_j^2$~\cite{Farr:2019rap,Talbot:2023pex}, much larger than the number of detected events, so the bias from this effect is negligible in our study.
}

We adopt a simple detection threshold requiring the combined SNR from Hanford and Livingston to be $\gtrsim 12$.
\footnote{As noted in \cite{Leyde:2023iof}, this is an approximation, as the application of this method to real data will require more intricate selection criteria, such as incorporating the false alarm rate.
}.
{
We use a two-layer SNR-threshold filter to quickly determine whether an event is observable.
First, we use a semi-analytic estimation formula (Note that this estimated value is not the SNR itself, but is approximately proportional to the SNR.):
\begin{equation}
    \rho_\text{rough} = \left(\frac{(1+z)\mathcal{M}_c}{30 M_\odot}\right)^{5/6} \left(\frac{400 \text{Mpc}}{D_L(z)}\right) \sqrt{F_+^2 \frac{(1+\cos^2\iota)^2}{4} + F_\times^2 \cos^2\iota}
\end{equation}
We use a conservative threshold of $\rho_\text{rough}>8.8$.
In the second-layer filter, we use \texttt{ml4gw}~\cite{2025JOSS...10.8836B}, which allows us to generate waveforms in batch.
We use the IMRPhenomD waveform template supported by \texttt{ml4gw}.
We then calculate the SNR using the fiducial noise PSD provided by DINGO and apply a threshold of 12 for filtering.
In a small-sample test, we evaluated both filtering layers and found that all samples passing the second-layer SNR threshold filter also pass the conservative semi-analytic threshold 8.8.
Therefore, in the actual data-generation process, we first apply the first-layer filter to all generated samples, and only those that pass the first layer are further screened by the second layer.
}

\begin{table}[!htb]
    \centering
    \caption{PBH and ABH population model parameters and their priors}
    \small
    \begin{tabular} {cccccc}
        \toprule
        Parameter &  & Unit & Prior \\
        \hline \hline
        $M_c$ & The characteristic mass & $M_\odot$ & U$(20,70)$  \\
        $\log_{10} \sigma$ & The width of the distribution & - & U$(-1,0)$  \\
        $\log_{10} f_{\rm PBH}$ & The fraction of PBHs in dark matter & - & U$(-5,0)$ \\
        \hline
        \hline
        $\log_{10} R_{\rm ABH}^0$ & Integrated merger rate of ABHs at $z=0$ & ${\rm Gpc}^{-3}{\rm yr}^{-1}$ & U$(0 ,3)$  \\
        $\log_{10} \lambda$ & Fraction of the Gaussian component in the primary mass distribution & - & U$(-6,0)$ \\
        $m_{\rm min}$ & Minimum mass of the power law component in the primary mass distribution & $M_\odot$ & U$(10,20)$ \\
        $m_{\rm max}$ & Maximum mass of the power law component in the primary mass distribution & $M_\odot$ & U$(40,100)$ \\
        $\alpha$ & Inverse of the slope of the primary mass distribution for the power law component & - & U$(-4,12)$ \\
        $\mu_G$ & Mean of the Gaussian component & $M_\odot$ & U$(20,50)$  \\
        $\sigma_G$ & Width of the Gaussian component & $M_\odot$ & U$(1,10)$ \\
        $\delta_m$ & Range of mass tapering on the lower end of the mass distribution & $M_\odot$ & U$(0,10)$  \\
        $\beta$ & Spectral index for the power law of the mass ratio distribution & - & U$(-4,12)$ \\
        \toprule
    \end{tabular}
    \label{tab:prior}
\end{table}

The single event likelihood is usually provided as posterior samples.
The integral appearing in \eqref{eq:tot_likelihood} for each event can be approximated using Monte-Carlo integration as \cite{LIGOScientific:2020kqk}
\begin{equation} \label{eq:event_MCint}
    \int\mathcal{L}(d_i|\boldsymbol{\theta})\pi(\boldsymbol{\theta}|\Lambda){\rm d}\boldsymbol{\theta}\approx
    \frac{1}{n_i}\sum_{j=1}^{n_i}\frac{\pi(\boldsymbol{\theta}_{ij}|\Lambda)}
    {\pi_{\varnothing}(\boldsymbol{\theta}_{ij})}\equiv\frac{1}{n_i}
    \sum_{j=1}^{n_i}\omega_{ij},
\end{equation}
where $\boldsymbol{\theta}_{ij}$ denotes the parameters of the
$j$-th sample ($n_i$ posterior samples in total) of the $i$-th event,
and $\pi_{\varnothing}(\boldsymbol{\theta}_{ij})$ is the prior used for the initial
parameter estimation. The sum above is taken over the posterior samples
$\boldsymbol{\theta}_{ij}\sim p(\boldsymbol{\theta}_{ij}|d_i)$.
{
Following Refs.~\cite{Farr:2019rap,Talbot:2023pex}, we define the effective sample size associated with the weighted Monte Carlo sum in \autoref{eq:event_MCint} as
\begin{equation}
    n_{i,{\rm eff}}(\Lambda)=\frac{\left(\sum_{j=1}^{n_i}\omega_{ij}\right)^2}{\sum_{j=1}^{n_i}\omega_{ij}^2}.
    \label{eq:event_ess}
\end{equation}
This quantity is the number of equally weighted posterior samples that would give the same Monte Carlo variance as the weighted estimate.
For each event, we used 10,000 posterior samples inferred by DINGO, which yield an average effective sample size of $\langle n_{i,{\rm eff}}\rangle>1000$.
}

Finally, the log-likelihood is evaluated as \cite{Mastrogiovanni:2023zbw}
\begin{equation}
    \ln \mathcal{L}(\boldsymbol{d}|\Lambda)\approx-\frac{T_{\rm obs}}{N_{\rm inj}}
    \sum_{j}^{N_{\rm det}}s_j + \sum_i^{N_{\rm obs}}\ln\left(\frac{T_{\rm obs}}{n_i}
    \sum_j^{n_i}\omega_{ij}\right).
\end{equation}
The posterior of the hyperparameters $\Lambda$ given the observed dataset $\boldsymbol{d}$ is obtained by \texttt{dynesty}~\cite{2020MNRAS.493.3132S}.

\subsection*{Neural network architecture}
\label{subsec:FA}

For the event-parameter vector $\theta$, we first standardize each parameter by subtracting its mean and dividing by its standard deviation.
Next, each event is encoded into a 31-dimensional summary via a 2-layer MLP with hidden‐unit sizes of [32, 32].
By merging the event axis into the batch axis, this transformation can be executed in parallel for all events.
Then, for each event's 31-dimensional vector, we append the observed number of events, $N_\text{obs}$, to create a 32-dimensional vector.
These vectors serve as the inputs to our Transformer encoder.

Within each Transformer encoder block, we first apply layer normalization, then perform multi-head self-attention (with 4 heads and scaled dot-product attention).
After adding a residual connection from the block’s input, we apply a second layer normalization and pass the result through a three-layer feedforward MLP (256 neurons per layer).
Finally, we add another residual connection from the self-attention output to produce the block’s output. Our model stacks 5 such Transformer encoder blocks in sequence.

The normalizing flow (NF) is responsible for reconstructing the posterior distribution.
It offers an effective method for representing complex probability distributions using neural networks. This approach facilitates efficient sampling and density estimation by expressing the distribution as a sequence of mappings, or "flows", $f : u \to \Lambda$, which maps from a simpler base distribution $u$ (a standard normal distribution $\mathcal{N}(0,1)$ in our case) to the parameter space (the hyperparameters of the population model in our case).
When the mapping $f$ depends on the observed data, denoted as $f_{\boldsymbol{d}}$, it describes a conditional probability distribution $q(\Lambda | \boldsymbol{d})$. The probability density function is given by the change of variables formula:
\begin{equation} \label{eq:A1}
    q(\Lambda|\boldsymbol{d})=\mathcal{N}(0,1)^D(f_{\boldsymbol{d}}^{-1}(\Lambda))|{\rm det}f_{\boldsymbol{d}}^{-1}(\Lambda)|,
\end{equation}
where $D$ is the dimensionality of the parameter space.

It is essential to design and combine multiple transformations to model complex distributions. Each transformation should be both invertible (so that $f_{\boldsymbol{d}}^{-1}(\Lambda)$ can be evaluated for any $\Lambda$) and possess a tractable Jacobian determinant (allowing for efficient computation of ${\rm det}f_{\boldsymbol{d}}^{-1}(\Lambda)$). These properties enable efficient sampling and density estimation, as described in equation \eqref{eq:A1}. Various normalizing flow architectures have been developed to satisfy these conditions, typically by composing several simpler transformations $f^{(j)}$, with each transformation being parameterized by the output of a neural network. To sample from the posterior, $\Lambda \sim q(\Lambda | \boldsymbol{d})$, we first sample $u \sim \mathcal{N}(0,1)^D$ and then apply the flow in the forward direction.

For each flow step, we employ a conditional coupling transformation \cite{2019arXiv190604032D}.
In each setup, the first $k$ components of the input are fixed, while the others undergo an elementwise transformation conditioned on all components and the data:
\begin{equation}
    f_{\boldsymbol{d},i}^{(j)}(u)=
    \begin{cases}
        u_i & \text{if } i\le k,
        \\
        f_i^{(j)}(u_{1:D},\boldsymbol{d}) &  \text{if } k< i\le D.
    \end{cases}
\end{equation}
When the elementwise functions $f_i^{(j)}$ are chosen to be monotonic, quadratic, rational spline functions, the transformation inherently satisfies the conditions required for a normalizing flow.
The spline parameters for each $f_i^{(j)}$ are determined by residual networks (ResNet), which takes as input the concatenated data, $u_{1:D}$ and $\boldsymbol{d}$.
To maintain the flexibility of the entire flow, we randomly permute the parameters between each transformation.
Each residual network consists of 4 hidden layers, with 16 neurons in each layer, and takes the output of the Transformer encoder as its input.
We employ a 9-step flow, meaning we utilize four residual neural networks.

\subsection*{Generating the data set}

{
We sample $\Lambda$ from its prior $\pi(\Lambda)$ using scrambled Sobol sampling.
For each $\Lambda$, we draw 6400 detectable events, characterized by $\boldsymbol{\theta} = \{m_1, m_2, d_L\}$ for each event, according to the population model for each sample $\Lambda$.
This is obtained by performing rejection sampling from the merger rate distribution based on the detection criteria mentioned above.
Meanwhile, for each $\Lambda_i$, we also use the same injection-recovery samples (\autoref{eq:injection}) as those used in the traditional HBA approach to compute the detectable event rate.
We also verified that directly estimating the detectable event rate from the acceptance fraction in the generation process yields consistent results, which further confirms that the number of injected samples ($2^{19}$) used to construct the injection-recovery set is sufficiently large to keep statistical errors small.

During each training epoch, for each hyperparameter $\Lambda$, we randomly take $n_\text{event} = 128$ out of the 6400 events we generated as a catalog.
These can help prevent overfitting.
Combined with the observation time $T_{\rm obs}$, which is dynamically sampled from a Poisson distribution whose expectation value is the recorded detectable event rate, our training dataset contains $(n_{\rm event} \times 3) + 1$ scalars for each $\Lambda$.
We generate $32768$ population samples (i.e., $\Lambda$).
}
We use 90\% of the samples for training and the remaining samples for testing.

\subsection*{Training the networks}

Fitting a flow-based model $q(\Lambda | \boldsymbol{d})$ to a target distribution $p(\Lambda | \boldsymbol{d})$
can be done by minimizing some divergence or discrepancy between them.
One of the most popular choices is the forward Kullback-Leibler (KL) divergence that is mass-covering \cite{Kullback:1951zyt},
\begin{equation}
    D_{\rm KL}(p||q)=\int{\rm d}\Lambda p(\Lambda|\boldsymbol{d})\ln\frac{p(\Lambda|\boldsymbol{d})}{q(\Lambda|\boldsymbol{d})}.
\end{equation}
This measure indicates how much information is lost when using $q$ as an approximation to $p$. The forward KL divergence is well-suited for situations in which we have samples
from the target distribution (or the ability to generate them), but we cannot necessarily
evaluate the target density. By taking the expectation over data samples $\boldsymbol{d} \sim p(\boldsymbol{d})$, we can simplify the expression, resulting in the loss function:
\begin{align}
    L&= \int {\rm d}\boldsymbol{d}p(\boldsymbol{d})\int{\rm d}\Lambda p(\Lambda|\boldsymbol{d})\ln\left(\frac{p(\Lambda|\boldsymbol{d})}{q(\Lambda|\boldsymbol{d})}\right) \notag \\
    &=\int {\rm d}\boldsymbol{d}p(\boldsymbol{d})\int{\rm d}\Lambda p(\Lambda|\boldsymbol{d})[-\ln q(\Lambda|\boldsymbol{d})]+\text{constant.} \notag \\
    &=\int{\rm d}\Lambda p(\Lambda)\int {\rm d}\boldsymbol{d}p(\boldsymbol{d}|\Lambda)[-\ln q(\Lambda|\boldsymbol{d})]+\text{constant.}
\end{align}
On the third line, we applied Bayes’ theorem to rewrite the cross-entropy between the two distributions in terms of the likelihood, thereby avoiding dependence on the unknown true posterior.
The posterior function we learn here is
\begin{equation}
    p(\Lambda | \boldsymbol{\theta}, N_\text{obs}) \propto \pi(\Lambda) N(\Lambda)^{N_\text{obs}} e^{-N(\Lambda)\xi(\Lambda)} \prod_{i=1}
    ^{N_{\rm obs}} \mathcal{L}(d_i|\boldsymbol{\theta}_i)\pi(\boldsymbol{\theta}_i|\Lambda),
\end{equation}
Because normalizing flows directly output normalized distributions, recovering the right-hand side of this expression requires multiplying by a normalization factor (the integral over $\Lambda$, i.e., the evidence).
Therefore, to recover results consistent with \autoref{eq:tot_likelihood} at inference time, we use as weights the normalization factor divided by the event parameter prior used in DINGO inference ($\pi_{\varnothing}$ in \autoref{eq:event_MCint}), perform weighted sampling of $\theta_{ij}$ from DINGO posterior samples, and then feed $\theta_{ij}$ together with $N_\text{obs}$ into the neural network to generate posterior samples of $\Lambda$. Collecting all sampled $\Lambda$ points reconstructs \autoref{eq:tot_likelihood}.
Since the trained neural network can perform batched inference over different $\theta_{ij}$, this procedure is computationally efficient.

Finally, the loss function can be approximated on a mini-batch of samples, omitting constant terms that are independent of the flow's parameters,
\begin{equation}
    L\approx-\frac{1}{N}\sum_{i=1}^{N}\ln q(\Lambda^{(i)}|\boldsymbol{d}^{(i)}),
\end{equation}
where $N$ samples are drawn ancestrally in a two-step process: sample from the prior, $\Lambda^{(i)}\sim p(\Lambda)$ and simulate data according to the population model, $\boldsymbol{d}^{(i)}\sim p(\boldsymbol{d}|\Lambda^{(i)})$.
Minimizing the above Monte Carlo approximation of the loss function is equivalent to
fitting the flow-based model to the samples by maximum likelihood estimation.

In order to avoid over-fitting, we add a $L^2$ regularization term to the loss function:
\begin{equation}
    L = -\frac{1}{N}\sum_{i=1}^{N}\ln q(\Lambda^{(i)}|\boldsymbol{d}^{(i)}) + \lambda \sum_\text{NN}{w},
\end{equation}
where $\sum_\text{NN}{w}$ is the sum of weights in the network and $\lambda$ is chosen to be $1 \times 10^{-3.5}$.
{
This hyperparameter was optimized jointly with the other hyperparameters using Optuna~\cite{Akiba:2019lwq}.
}
We then take the gradient of $L$ with respect to network parameters and minimize using the AdamW optimizer \cite{Loshchilov:2017bsp}.
The learning rate is started from $10^{-3}$ and is reduced during training using the Plateau scheduler.

In our work, the parameters of all networks are optimized jointly with a batch size of 128.

\begin{figure}[!htb]
    \centering
    \includegraphics[width=0.5\textwidth]{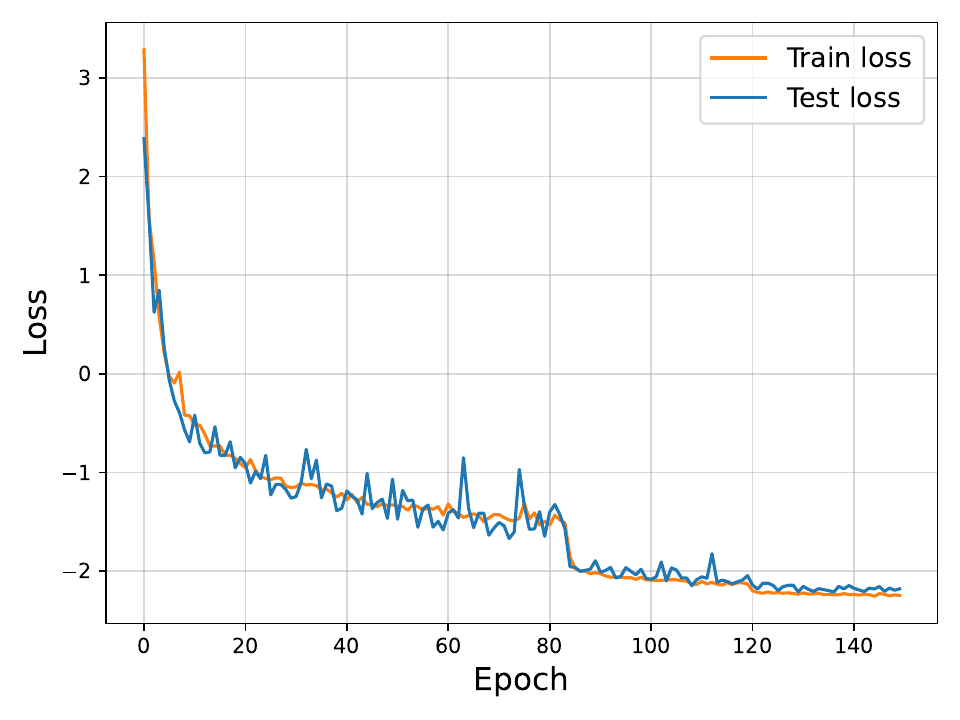}
    \caption{\textbf{Loss decay during training.}
    Since the test loss (blue) and train loss (orange)
    do not differ much, we conclude that the model can generalize effectively to data that were not included in the optimization process.}
    \label{fig:loss_curve}
\end{figure}

We trained the model for 150 epochs, although the training and test loss curves shown in \autoref{fig:loss_curve} indicate that convergence is already achieved by 120 epochs.
The train and test losses coincide with only minor differences, suggesting that the model can process unseen input data and generate accurate hyperparameter posterior distributions.
{
To clarify the role of the Transformer encoder in our network architecture, we also compared against an architecture that uses ResNet without the Transformer encoder.
For a fair comparison, we also used Optuna to tune the ResNet hyperparameters, such as the number of layers and the number of neurons per layer, and selected the best-performing model.
For the ResNet architecture, the best model achieves a test loss of $-1.84$, whereas the architecture with the Transformer encoder achieves $-2.24$, indicating that the Transformer encoder indeed improves our network performance.
}

\subsection*{Performance under extreme scenarios}

\begin{figure}[!htb]
    \centering
   \includegraphics[width=0.48\textwidth]{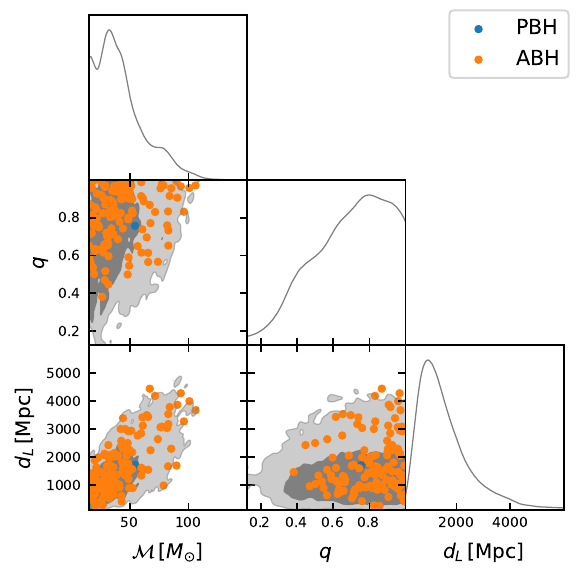}
   \includegraphics[width=0.48\textwidth]{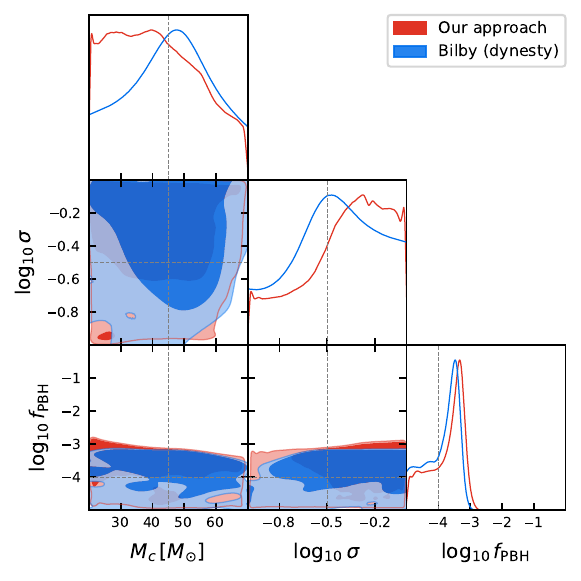}
\caption{\textbf{\textit{Left}: Mock observation event parameter
distribution.} Gray contours show joint posterior of chirp mass
$\mathcal{M}$, mass ratio $q$, and luminosity distance $d_L$
inferred by DINGO for all events, which serve as input to
our population analysis.
    Blue points mark the true parameters of individual PBH merger events, while orange points mark the true parameters of individual ABH merger events.
    \textbf{\textit{Right}: Hyperparameter inference for PBHs in
a black hole population model composed of PBHs and ABHs
($f_\text{PBH} = 10^{-4}$).}
    The red regions show the inference results from our approach, while the blue regions show the results from Hierarchical Bayesian inference using dynesty for sampling.
    The injected values are indicated by dashed lines.
    In both panels, the contours on the two-dimensional plane represent the 68\% and 95\% confidence regions.}
    \label{fig:dist_extreme}
\end{figure}

In this subsection, we examine performance in ABH-dominated regimes by fixing $f_{\rm PBH}$ to $10^{-4}$.
The remaining population hyperparameters are the same as in the scenario tested above.
In this case, only one event is a PBH binary.
\autoref{fig:dist_extreme} displays the inferred population hyperparameter posteriors.
In this scenario, both our method and the traditional method recover the posterior peak of $f_\text{PBH}$.
Similar to the case with $f_{\rm PBH} = 10^{-3}$, our inferred posterior is comparatively more conservative.
Using dynesty, we obtain a $2\sigma$ upper limit of $\log_{10} f_{\rm PBH} < -3.10$.
For our approach, we obtain $\log_{10} f_{\rm PBH} < -3.26$.
The injected value lies within the upper limits inferred by both methods.

\bibliography{sample}

\section*{Acknowledgements}

This work is supported by National Key Research and Development
Program of China, No. 2021YFC2203004, and NSFC, No. 12075246. We
acknowledge the use of high performance computing services
provided by the International Centre for Theoretical Physics
Asia-Pacific cluster, the Tianhe-2 supercomputer and Scientific
Computing Center of University of Chinese Academy of Sciences.

\end{document}